# PLANAR VACUUM DIODE WITH MONOENERGETIC ELECTRONS

**Dimitar G. Stoyanov**

Department of Physics, Sofia Technical University, Sliven Branch

59, Bourgasko Shaussee Blvd, 8800 Sliven, Republic of Bulgaria

E-mail: dstoyanov@tu-sliven.com

**Abstract**

*The paper analyses volt-ampere characteristics of a planar vacuum diode with mono-energetic electrons, emitted by the cathode (an electron beam). The movement of the electron beam in the volume of the diode is described. An analytic dependence of the volt-ampere characteristics in an unlimited and limited by the field mode of the electron beam is derived.*

## 1. INTRODUCTION

The development of physics gains great significance in relation to the important investigations of the physical processes carried out alongside with the assembly of actively interacting particles. In reality there are such assemblies in the plasma-chemical processes as well as in the intensive electron and ion beams of accelerators, ion engines and the lasers of free electrons. For example, the electron cloud formed in the vacuum tubes represents such an assembly.

The theory of the planar vacuum diode with thermoionic emission from the cathode electrons was worked out in the beginning of XX c., when the first vacuum tubes were developed. Basic works in this respect belong to Child [1], Langmuir [2, 3] and are elaborated by others [4, 5]. These studies suggest that the planar diode is a system of two planar infinite metal electrodes at a distance of **d** from each other set in a vacuum glass tube (absolute permitivity of vacuum is **ε₀)**. The cathode emits electrons (electron mass is **m$_e$** and electronic electric charge is **q=-e** (negative)) as a result of the thermoionic emission. The electrons emitted by the cathode fall in the volume between the two electrodes and if there is an electric field formed by the potential difference **U** between the two electrodes, an electric current with density of **j** flows through the diode [2, 3].

$$j = \frac{4.\varepsilon_0}{9} \cdot \sqrt{\frac{2.e}{m_e}} \cdot \frac{U^{3/2}}{d^2} \qquad (1)$$

This is the well-known Child–Langmuire's law. No matter how correctly the law describes the qualitative aspect of the dependence of the current through the diode as a function of applied exterior voltage; there have constantly been attempts to modify Child–Langmuire's law in order to use it in relation to specific problems. The results, however, have been partial and lack general theoretical conclusions.

The following discrepancies based on Child–Langmuire's law alongside with vacuum tube tests can be pointed out and some hot questions can be discussed as well:

**1.** The cathode of the real vacuum diode in stationary mode is heated up to constant temperature of **T** and emits electrons as a result of the thermoionic emission. The frequency of electron emission is constant and is controlled by the temperature of the cathode. The density of electronic current emitted by the cathode is defined by Richardson-Dushman's formula [6]. Thus, the frequency of electron emission by the cathode depends mainly on temperature but on intensity of the electric field, the latter has no such great importance.

a) According to dependence (1) current depends on voltage, not on the cathode temperature. Taking general considerations into account, current should depend on the heating degree of the cathode.

b) According to dependence (1) current is a monotonous rising function of voltage on the diode and can assume unlimited values. According to Richardson-Dushman's law current density through the diode cannot exceed current density of the electrons emitted by the cathode. The maximum value of the current through the diode is called saturation current, the latter depending not so much on voltage.

c) If **U=0 V**, according to (1), there won't be any current through the diode. A simple measurement shows that there is current, no matter how small it is (from the order of micro-amperes);

d) According to dependence (1) the change of voltage determines current density through the diode. At the same time the cathode emits electrons with one and the same frequency. Where do some of the emitted electrons, especially at low voltages, go?

**2.** Dependence (1) does not include quantities connected with the type of the electron distribution function of emitted electrons in speed – is distribution function Maxwellian or non-Maxwellian?

**3.** Dependence (1) does not state if it is valid for vacuum photocell working on the principle of photoelectric effect lit by light with a certain spectral distribution.

The above mentioned issues require a more detailed approach to the dependence of the current through the diode on voltage between the two electrodes, taking into consideration the characteristics and functions of the distribution of the emitted by the cathode electrons. The results should be valid both for the planar vacuum diode as well as planar vacuum photocell.

The present paper is limited to the planar vacuum diode in stationary mode. To simplify matters, we shall analyse the case when electrons with equal initial speed (mono-energetic electrons) are emitted by the cathode.

## 2. DIODE WITH MONOENERGETIC ELECTRONS

### 2.1. GEOMETRY AND INITIAL EQUATIONS

We assume that the diode is a system of two infinite parallel metal plane at a distance of **d**, set in high vacuum. The space symmetry is such that we can choose Cartesian coordinates, where the axes **OX** and **OY** lie on the metal plane which is a cathode, while axis **OZ** is perpendicular to the cathode. Thus, all units describing the diode parameters depend only on **z**. The cathode area has a **z** coordinate: **z=0**, whereas the anode area has a coordinate **z=d**.

In the case under discussion the cathode emits electrons with identical kinetic energy $E_0$ (mono-energetic electrons). They move at a speed of $v_e(z)$ in the electric field existing between the cathode and the anode with potential $\varphi(z)$. There is a potential difference of **U** between the cathode and the anode. We assume that the cathode electric potential is zero: $\varphi(z=0)=0.$ Then the potential of the field on the anode is: $\varphi(z=d)=U.$

The speed of electron motion $v_e(z)$ can be expressed by the potential of the field [3,5]

$$v_e(z) = \pm\sqrt{\frac{2.(E_0 - q.\varphi(z))}{m_e}}. \tag{2}$$

The current density in the volume between the anode and the cathode $j(z)$ formed by the moving mono-energetic electrons in the case of the planar stationary diode does not depend on the coordinate $z$ as a consequence of the law for preservation of electrical charge. The magnitude of $j$ depends on the mode of the working diode. No matter what the mode of the working diode is, the density of the emitted electrons has a magnitude of $j_0$.

According to [1-5] the potential of the electrical field $\varphi(z)$ in the volume between the anode and the cathode can be connected with the density of the electrons $n_e(z)$ using Poisson's equation in the planar case.

$$\frac{d^2\varphi(z)}{dz^2} = -\frac{q.n_e(z)}{\varepsilon_0}. \tag{3}$$

Besides, according to [1-5] the electron density can be expressed by the current density through the diode and the speed of the electrons defined in equation (2):

$$n_e(z) = \frac{j}{q.v_e(z)} = \frac{j}{q.\sqrt{\frac{2.(E_0 - q.\varphi(z))}{m_e}}}. \tag{4}$$

Next discussion does need specification of current density values of the current flowing through the diode. That is the reason why we are investigating two modes of the working diode. The simple case comes first.

## 2.2 MODE OF UNLIMITED ELECTRON BEAM

In that mode every emitted by the cathode electron reaches the anode so that the only electrons in the volume are the ones moving from the cathode to the anode. Therefore, the current density is equal to the current density of the emitted by the cathode electrons **$j_0$** and it should be expected that the current flowing through the diode does not depend on the potential difference between the electrodes. There is not an analogy of the above statement in all works or literature.

If equation (4) is replaced in equation (3) taking into consideration the magnitude of the current, we get unlinear second-order ordinary differential equation from for defining the electrical field potential in the diode volume in that working mode.

$$\frac{d^2\varphi(z)}{dz^2} = -\frac{j_0}{\varepsilon_0 \cdot \sqrt{\frac{2 \cdot (E_0 - q \cdot \varphi(z))}{m_e}}}. \tag{5}$$

After a change of the variables into dimensionless units the differential equation looks like this [5].

$$\frac{d^2\Phi}{dx^2} = \frac{J_0}{\sqrt{1+\Phi}}, \tag{6}$$

where: - $x = \frac{z}{d}$ stand for dimensionless coordinates;

- $\Phi(x) = -\frac{q \cdot \varphi(z)}{E_0}$ is a dimensionless potential of the electric field;

- $J_0 = \frac{q \cdot j_0 \cdot d^2}{\varepsilon_0 \cdot E_0} \cdot \sqrt{\frac{m_e}{2 \cdot E_0}}$, $J_0 > 0.$ stands for dimensionless current density

First integration of equation (6) gives us equation (7), as well as in [5], where it is accepted that the integration constant standing on the right of (7) is zero.

$$4 \cdot J_0 \cdot \sqrt{1+\Phi} - \left(\frac{d\Phi}{dx}\right)^2 = \text{const.} \tag{7}$$

We accept, however, that the constant is positive. Firstly, it is proportional to the speed of the electrons in an area where the intensity of the electrical field is zero. Secondly, as we have accepted earlier, the speed of the electrons is always a positive value in this specific working mode (i.e. there is no space point where it is zero). Then in the point with zero intensity the potential of the field $\Phi$ has a minimum value of $\Phi_m$ (according to equation (6) the second derivative is positive). The coordinate of this point is marked with $x_m$ - position of the minimum of the field potential.

Therefore, we consider equation (7) as follows:

$$4.J_0.\sqrt{1+\Phi} - \left(\frac{d\Phi}{dx}\right)^2 = 4.J_0.\sqrt{1+\Phi_m}. \tag{8}$$

The solution to the equation (8), which is the second integration in the case, can be expressed in the following way (9):

$$\frac{2}{3}\cdot\frac{1}{\sqrt{J_0}}\cdot\sqrt{\sqrt{1+\Phi} - \sqrt{1+\Phi_m}}\cdot\left(\sqrt{1+\Phi} + 2.\sqrt{1+\Phi_m}\right) = \pm(x - x_m). \tag{9}$$

The sign (-) in equation (9) corresponds to the case $x \leq x_m$, whereas the sign (+) stands for the case of $x \geq x_m$. A graph example of the dependence (9) is given in Figure 1.

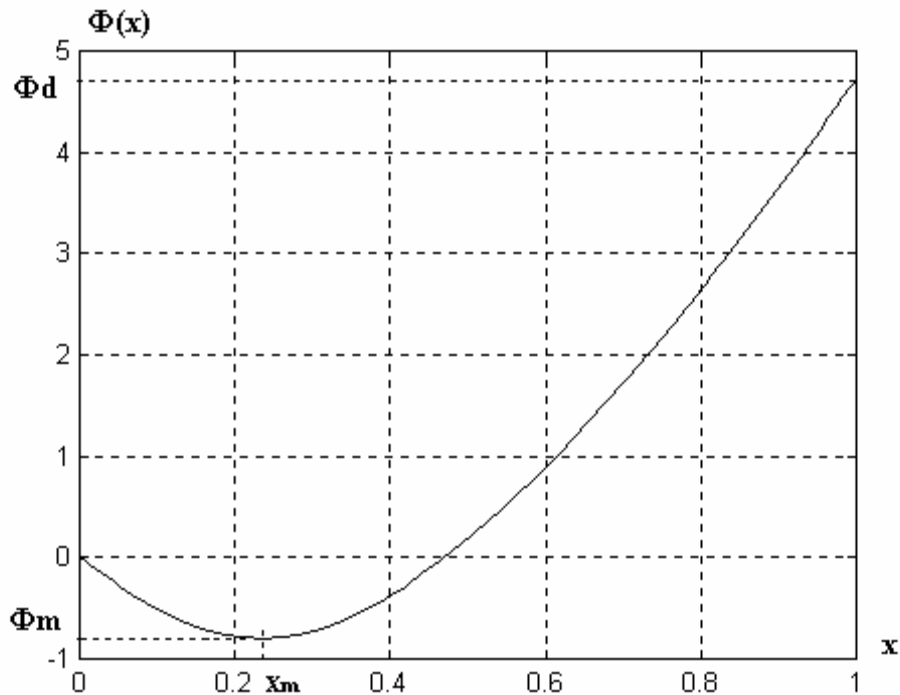

**Fig. 1** Dependence of the electrical field potential on the x coordinate when $J_0 = 16$ and $\Phi_m = -0.8$.

The figure shows that the further the cathode is, the initial potential of the electrical field becomes less. Then it reaches a local minimum and it starts to go up so as when it reaches the anode it has a value of $\Phi(x=1) = \Phi_d$.

From physical point of view, all electrons from the beam fall on the anode only when $\Phi_m > -1$ (this is an energetic variant of the condition of unlimited beam electrons and it is a consequence of the requirement for the electron speed in the minimum to be higher than zero).

Credit should be paid to Shottky [4] for the definition that a potential minimum $\Phi_m$ exists and it limits the current. The magnitude of $\Phi_m$ and $x_m$ act as integration constants of (5) in solution to the equation (6).

If in equation (9) we replace the coordinates and the magnitude of the potential separately on the cathode $\Phi(x=0) = 0$ and the anode $\Phi(x=1) = \Phi_d$, using term-by-term addition we can obtain the connection between the current density and the voltage on the vacuum diode with an electron beam.

$$1 = \frac{2}{3} \cdot \frac{1}{\sqrt{J_0}} \cdot \left[ \sqrt{\sqrt{1+\Phi_d} - \sqrt{1+\Phi_m}} \cdot \left(\sqrt{1+\Phi_d} + 2\cdot\sqrt{1+\Phi_m}\right) + \right.$$
$$\left. + \sqrt{1 - \sqrt{1+\Phi_m}} \cdot \left(1 + 2\cdot\sqrt{1+\Phi_m}\right) \right] \qquad (10)$$

The first part of the sum on the right of the equation is the distance between the anode and the potential minimum, while the second part of the sum is the distance between the potential minimum to the cathode. The two values are from dimensionless kind. Their sum is the distance between the anode and the cathode **d**, i.e. if expressed in dimensionless kind it is 1, which is the left side of the equation.

The analysis of the obtained solution (10) shows that if $J_0$ is defined, equation (10) can be regarded as an equation for $\Phi_d(\Phi_m)$. That solution exists at a wide range of magnitude for $\Phi_m$ (taking into consideration the condition of unlimited beam). The values of $\Phi_d$ have a wide range of magnitude to and the dependence of $\Phi_d$ on $\Phi_m$ is ambiguous. Figure 2 shows the dependence of $\Phi_d(\Phi_m)$ in a graph.

What conclusions can be made about the investigated unlimited mode of electrons? As it has been observed, there are multiple values of potential difference $\Phi_d$ between the electrodes when current flows through the diode at fixed $J_0$.

The explanation can been that the volt-ampere characteristics of the diode in the mode of unlimited electron beam looks like the horizontal line 3 in Fig. 3, i. e. constant in magnitude current flows through the diode if the voltage of the diode has bigger or equal magnitude to the defined minimum value.

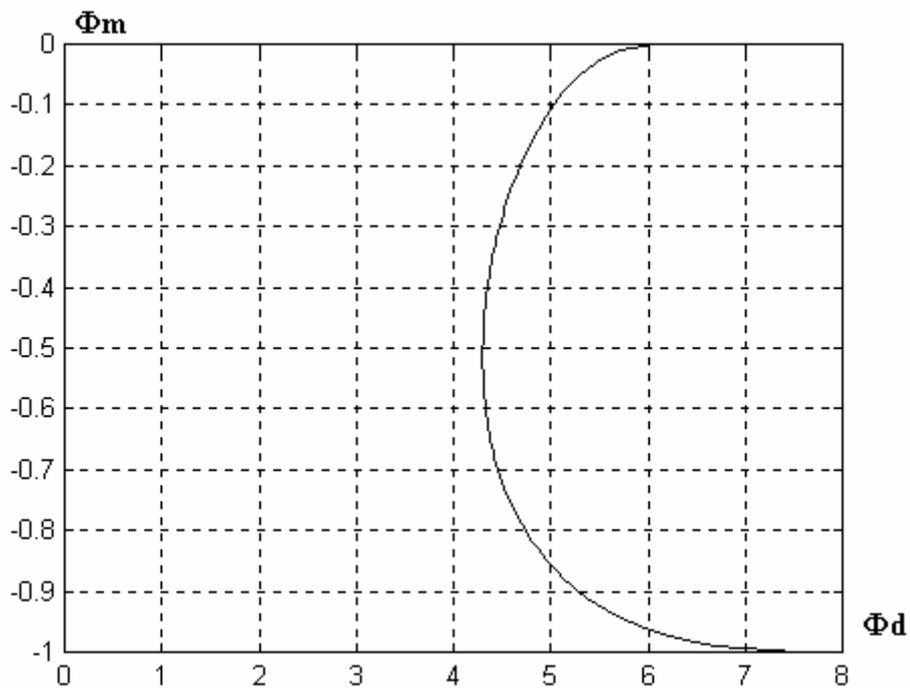

**Fig. 2** The dependence between $\Phi_d$ and $\Phi_m$ when $J_0 =16$ and the mode is that of unlimited beam.

This volt-ampere characteristic in itself corresponds to the saturation section of the vacuum diode or to the planar vacuum photocell.

### 2.3 MODE OF LIMITED ELECTRON BEAM

We consider that in this working mode of the diode the cathode emits electrons with current density $j_0$ and they start moving from the cathode to the anode. When the electrons reach the minimum of the electrical field potential at $z_m$ they have zero speed according to our assumptions. It is necessary for them to stop so that some of them go back to the cathode which absorbs them whereas the rest continue to the anode.

The electrons going to the anode form current density of **j** (**j** < **j₀**) in the area of $z > z_m$. The electrons coming back to the cathode in the area of $z < z_m$ possess the same speed in magnitude and so do the moving electrons from the cathode to the minimum position but in reverse, i. e. to the cathode. Thus, in this area the sum density of the current is again **j**, but it is difference of **j₀** (from the going to the minimum electrons) and **j₀ − j** (from the coming back to the cathode electrons). In this area, however, the density of the electrons is summed up. So the analogue of (6) in this working mode, we assume, is two different equations. The first is about the area from the cathode to the potential minimum and the second is about the area from the potential minimum to anode:

$$\frac{d^2\Phi}{dx^2} = \frac{2.J_0 - J}{\sqrt{1+\Phi}}, \qquad 0 \leq x \leq x_m. \tag{11a}$$

$$\frac{d^2\Phi}{dx^2} = \frac{J}{\sqrt{1+\Phi}}, \qquad x_m \leq x \leq 1, \tag{11b}$$

where:- $J = \frac{q.j.d^2}{\varepsilon_0.E_0} \cdot \sqrt{\frac{m_e}{2.E_0}}$, $J_0 > J > 0$. is dimensionless current though the diode.

The above obtained equations (11a) and (11b) are solved and the received solutions are adapted to $x_m$. The fact that the electron speed in the position of the minimum potential $x_m$ is zero shows that the potential has a value of $\Phi_m = -1$. This leads to the conclusion that the integration constant of (11a) and (11b) in their first integration (the analogues to (7)) is equal to zero.

Thus, we obtain:

$$4.(2J_0 - J).\sqrt{1+\Phi} - \left(\frac{d\Phi}{dx}\right)^2 = 0. \tag{12a}$$

$$4.J.\sqrt{1+\Phi} - \left(\frac{d\Phi}{dx}\right)^2 = 0. \tag{12b}$$

In the second integration of (11) we obtain

$$\frac{2}{3} \cdot \frac{1}{\sqrt{2.J_0 - J}} \cdot (1 + \Phi(x))^{3/4} = -(x - x_m), \qquad 0 \leq x \leq x_m. \tag{13a}$$

$$\frac{2}{3} \cdot \frac{1}{\sqrt{J}} \cdot (1 + \Phi(x))^{3/4} = +(x - x_m), \qquad x_m \leq x \leq 1. \tag{13b}$$

The solution shows that the curve of the potential is not symmetrical to the minimum. In the area to the minimum the field potential decreases faster in comparison with the mode of the unlimited beam, whereas in the area after the minimum the potential goes up more slowly.

If in (13a) and (13b), we replace the coordinates and the field potential of the cathode and the anode using term-by-term summing up we shall obtain the following volt-ampere diode characteristics in the limited mode.

$$1 = \frac{2}{3} \cdot \frac{(1 + \Phi_d)^{3/4}}{\sqrt{J}} + \frac{2}{3} \frac{1}{\sqrt{2.J_0 - J}}, \qquad J_0 > J > 0. \tag{14}$$

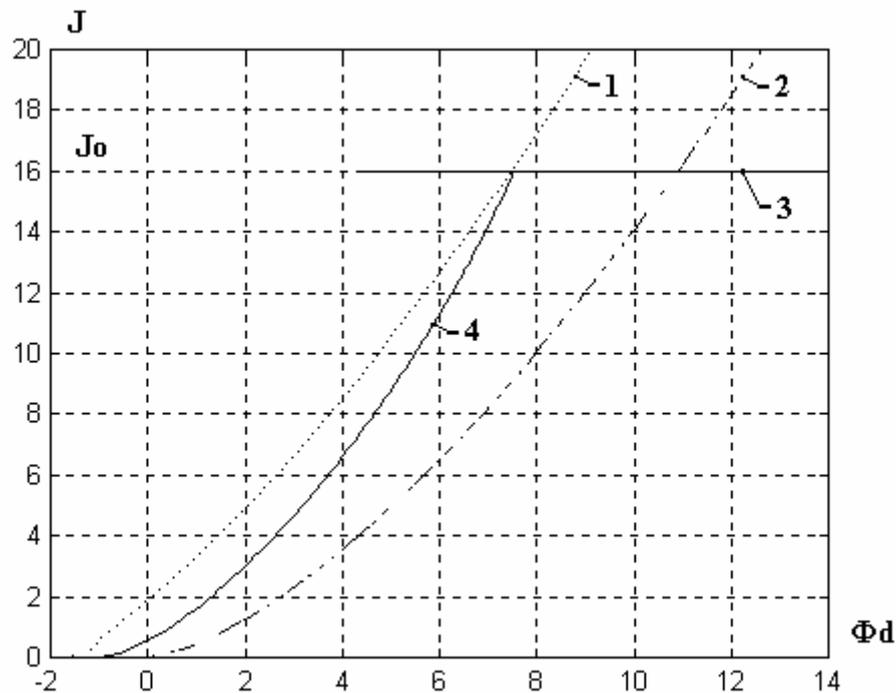

**Fig. 3** Volt-ampere characteristics of the diode when $J_0 = 16$: curve 1 - according to (16); curve 2 - according to (17); curve 3 - according to (10); curve 4 - according to (14).

Here as well as in (10), the first part of the sum on the right of the equation (14) is the distance between the anode and the potential minimum, while the second part of the sum is the distance between the potential minimum and the cathode. The two values are from dimensionless kind.

The graph of our solution (14) is given with line 4 in fig. 3.

The solution describes the rising part of the volt-ampere characteristics of the planar vacuum diode. According to equation (14) in the mode of limited beam at the change of the current from zero to saturation current $J_0$, $\Phi_d$ accepts values in the interval $[-1, \Phi_h]$, where:

$$\Phi_h = \left(\frac{3.\sqrt{J_0}}{2} - 1\right)^{4/3} - 1. \tag{15}$$

Current does not flow through the diode at $\Phi_d \leq -1$, when the value of $\Phi_d = -1$ reminds and coincides with the so-called 'stopping potential' of the volt-ampere characteristics of the vacuum photoelectric cell.

Another positive fact is that non-zero current flows through the diode at $\Phi_d = 0$.

The graph represents that the current through the diode in mode of limited beam goes up monotonously from zero to the value of saturation current, not being able to acquire higher values.

## 2.4 DISCUSSION OF RESULTS

In [5] the assignment suggests that the emitted electrons are with equal initial speed different from zero. We assume that $\Phi_m = -1$. The abovementioned points out that the current through the diode is limited and it does not record the coming back towards the cathode electrons. Thus, we can get from equation (10) the following

$$J_0 = \frac{4}{9} \cdot \left[(1+\Phi_d)^{3/4} + 1\right]^2. \tag{16}$$

The same formula we get in [5] but the current through the diode is on the left, whereas we consider it as saturation current. That is the reason why the equation in (16) corresponds to the point of volt-ampere characteristics of the diode (Fig. 3), where the vacuum diode goes through the mode of limited electron beam to the mode of unlimited electron beam. This happens when there is potential difference between the two electrodes $\Phi_d = \Phi_h$, which can be defined with equation (15). In the case of **$J_0$=16** we calculated that $\Phi_h = 7.5499$. The dependence graph (16) is given for comparison in Fig. 3 with dotted line.

If we turn back to equation (1) and its conclusion [1, 2], it suggests that $\Phi_m = 0$ and the electrons are emitted with initial speed of zero. Taking the two abovementioned assumptions into consideration in equation (8), we get.

$$J_0 = \frac{4}{9} \cdot \Phi_d^{3/2}. \qquad (17)$$

The same equation coincides with Child-Langmuire's law (1) (in dimensionless kind) but only on the right of the equation. We think there is saturation current on the left and not like in (1) unsaturated current. In the case of **$J_0$=16** we calculated from (17) the following: $\Phi_d = 10.9027$. The graph of equation (17) is given in Fig. 3 with dotted line for comparison.

If in [3] it suggested that $\Phi_m = 0$ and the electrons are emitted by the cathode at an initial speed of zero, then according to our assumptions, we get equation (8).

$$J_0 = \frac{4}{9} \cdot \left(\sqrt{1+\Phi_d} - 1\right) \cdot \left(\sqrt{1+\Phi_d} + 2\right)^2. \qquad (18)$$

The above equation is not a modification of Child-Langmuire's law in dimensionless kind but the saturation current on the left of equation (18) is not unsaturated current. Therefore, the expression (18) corresponds to the point of volt-ampere characteristics of the diode (Fig. 3) where the vacuum diode has a simple minimum potential of $\Phi_m = 0$ (ref. Fig. 2). In the case of **$J_0$=16** we calculated from (18) the following: $\Phi_d = 6.069$.

We shall repeat again that in works [1-3] the electrons are emitted by a heated cathode which suggests a presence of distribution of emitted electrons in speed and not of mono-energetic electrons. Therefore, the obtained dependencies can be accepted as a solution to the problem but only in the sense of asymptotic dependencies for great values of voltage.

Thus, we come to the statement that in works [1-5] we have a situation with dependencies obtained from non-argumentative values of initial speed of the emitted electrons without recording the coming back towards the cathode electrons and without any solid argumentation for values of $\Phi_m$ from physical point of view.

Fig. 4 depicts the obtained volt-ampere characteristics of the planar vacuum diode with mono-energetic electrons with one and the same initial speed but with different values of the saturation current. It is clear from the figure that the curves of the current rising for different values of the saturation current do not overlap.

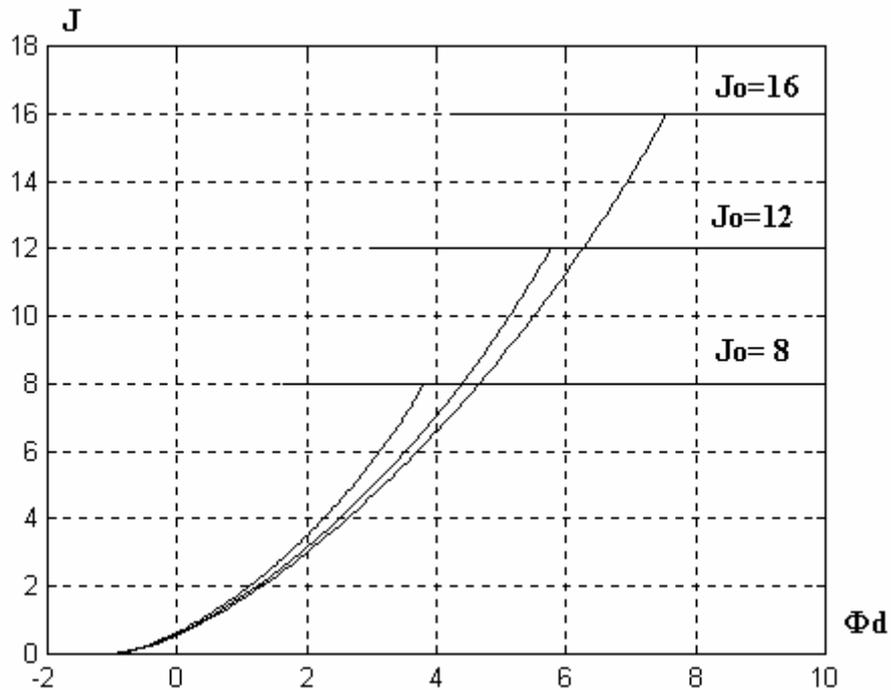

**Fig. 4** Volt-ampere characteristics of a planar vacuum diode
for different values of the saturation current.

## 3. CONCLUSION

To sum it up, an analytical form of the volt-ampere characteristics (14) of a planar vacuum diode with electron beam from mono-energetic electrons has been obtained. It has been clearly stated that the work of the diode in a limited mode by the field potential beam forms an area with rising current through the diode as long as it reaches the saturation current. The work of the diode in a mode of unlimited beam forms an area with the saturation current through the diode and the saturation current coincides with the emitted by the cathode current. Fig. 2 shows that the work of the diode in a mode of unlimited beam the volt-ampere characteristics has S - shaped fluctuation.